\documentclass[12pt]{article}
\pdfoutput = 1
\usepackage[colorlinks = true]{hyperref}
\hypersetup{citecolor = blue,
			linkcolor = blue,
			 urlcolor = blue}
\usepackage{amsthm, amsmath, amsfonts, amssymb}
\usepackage{graphicx}
\usepackage{natbib}
\usepackage{bm}

\newtheorem{theorem}{Theorem}
\newtheorem{corollary}{Corollary}
\theoremstyle{definition}

\newtheorem{remark}{Remark}

\usepackage{sectsty}
\allsectionsfont{\normalsize\bfseries}
\usepackage{caption}
\usepackage[normalem]{ulem}
\DeclareMathOperator*{\argmax}{argmax}

\newcommand{\blind}{0}

\begin{document}

\def\spacingset#1{\renewcommand{\baselinestretch}%
{#1}\small\normalsize} \spacingset{1}

%%%%%%%%%%%%%%%%%%%%%%%%%%%%%%%%%%%%%%%%%%%%%%%%%%%%%%%%%%%%%%%%%%%%%%%%%%%%%%

\if0\blind
{
  \title{\bf Taylor's Theorem and Mean Value Theorem for Random Functions and Random Variables}
  \author{Yifan Yang 
  	\thanks{
  		Emails: yiorfun@case.edu; xzhou126@umd.edu; mxw827@case.edu; 
  		and * is for the corresponding authors
  	} \hspace{.2cm}  \\
    Department of Population and Quantitative Health Sciences, \\
    Case Western Reserve University, Cleveland, OH \\
    and \\
    Xiaoyu Zhou \\
    Laboratory of Cognition \& Emotion, \\
    University of Maryland, College Park, MD \\
    and \\
    Ming Wang \footnotemark[1] \\
    Department of Population and Quantitative Health Sciences, \\
    Case Western Reserve University, Cleveland, OH
    }
  \maketitle
} \fi

\if1\blind
{
  \bigskip
  \bigskip
  \bigskip
  \begin{center}
    {\LARGE\bf Taylor's Theorem and Mean Value Theorem for Random Functions and Random Variables}
\end{center}
  \medskip
} \fi

\bigskip
\begin{abstract}
	This study addresses the often-overlooked issue of measurability at intermediate points when applying Taylor's theorems to random functions and random vectors (e.g., likelihood functions with respect to estimators) in statistics.
	Classical Taylor-related theorems were originally developed for deterministic settings.
	Consequently,  
	they do not directly extend to stochastic functions and variables and do not inherently guarantee the measurability of intermediate points.
 	In statistical contexts, applying these theorems without properly accounting for randomness can lead to analyses that lack well-defined probabilistic interpretations.
	Elementary approaches, such as pointwise constructions, are insufficient for handling random quantities and establishing measurable intermediate points. 
	Moreover, some statistical literature has implicitly disregarded this issue, often neglecting the stochastic nature of the problem and assuming that intermediate points are measurable.
	To address this gap, we develop multivariate Taylor's and mean value theorems tailored for random functions and random variables under mild assumptions.
	We provide illustrative examples demonstrating the applicability of our results to commonly used statistical methods, 
	including maximum likelihood estimation, $M$-estimation, and profile estimation. 
	%
	%Finally, 
	%we extend our results to cases where the highest-order derivatives exhibit quasi-continuity with closed graphs.
	Our findings contribute a rigorous foundation for the applications of Taylor expansions in statistics.
\end{abstract}

\noindent%
\vfill

\newpage
\spacingset{1.75} % DON'T change the spacing!

\section{Introduction} \label{Sec:introduction}

	Taylor-related theorems, including Taylor's theorem, Taylor series, and the mean value theorem (i.e., Taylor's theorem for the first-order derivative), are fundamental tools in calculus for analyzing the local properties of functions.
	Taylor polynomials provide linear approximations of complex functions, as originally formulated by \citet{Taylor1717}.
	Classical versions of these theorems, along with their extensions to Banach spaces, are detailed in \citet{Lang1993}.  
	For clarity and ease of comparison, 
	we present a classical version of the mean value theorem below:
	
	\textit{
		(Deterministic mean value theorem) 
		Suppose $\bm{x}$ and $\bm{y}$ are deterministic vectors, and  $f$ is a deterministic function satisfying certain smoothness conditions.
		Then, there exists a deterministic vector $\bm{\xi}$ such that $f\left(\bm{x} + \bm{y} \right) = f \left(\bm{x}\right) + \nabla f \left(\bm{\xi}\right) \bm{y}$, 
		where $\nabla f$ is the row vector of the gradient of $f$.
	}
	
	\noindent It is important to emphasize that the above classical mean value theorem (and the classical Taylor's theorem) follows a ``\emph{deterministic in, deterministic out}'' principle: 
	it requires $\bm{x}$, $\bm{y}$, and $f$ to be deterministic and yields a deterministic intermediate point $\bm{\xi}$.
	
	In statistics, 
	the mean value theorem and Taylor's theorem are widely used to establish important results, 
	such as the Delta method and the asymptotic properties of likelihood-based estimators.
	However, in this context, both functions and variables are typically constructed from the random sample,
	making them inherently random rather than deterministic. 
	Consequently, directly applying classical Taylor's theorems to these random functions and variables can be problematic and may lack mathematical rigor due to neglecting considerations of randomness.
	
	To illustrate this issue, 
	consider an example in establishing the asymptotic normality of the sample mean estimator $\widehat{\theta}$ based on a single observation $X \sim \operatorname{Poisson}(\theta)$, 
	where the population parameter $\theta > 0$ is unknown.
	The probability mass function of $X$ is $p(x; \theta) = \theta ^ {x} \exp(\theta) / x!$, 
	resulting in the log-likelihood function (for a single observation): $\ell(\theta; x) = - \theta + x \log(\theta) - \log(x!)$.
	The score function with respect to $\theta$ is given by $\ell ^ {(1)}(\theta; x) = - 1 + x / \theta$ and its derivative is $\ell ^ {(2)}(\theta; x) = - x / \theta ^ 2$.
	To explore the asymptotic properties of $\widehat{\theta}$,
	we replace the realization $x$ with the random variable $X$ in the log-likelihood function and consider the second-order Taylor expansion of the random log-likelihood function $\ell(\cdot; X)$, evaluated at $\widehat{\theta}$, around the true value $\theta$:
	\begin{equation*}
		\ell \left(\widehat{\theta}; X\right)
		= \ell \left(\theta; X\right)
		+ \left(\widehat{\theta} - \theta \right) \ell ^ {(1)}(\theta; X)
		+ \frac{1}{2} \left(\widehat{\theta} - \theta \right) ^ 2 
		\ell ^ {(2)}\left(\widetilde{\theta}; X\right), 
	\end{equation*} 
	where $\widetilde{\theta}$ is a ``point'' between $\widehat{\theta}$ and $\theta$.
	Two key questions arise regarding the above application of the classical Taylor's theorem: 
	
	(1) \emph{Does the classical Taylor's theorem remain valid for the random function $\ell(\widehat{\theta}; X) = - \widehat{\theta} + X \log(\widehat{\theta}) - \log(X!)$ and the random variable $\widehat{\theta} \triangleq \widehat{\theta}(X)$}?
	
	(2) \emph{Can the classical Taylor's theorem justify that the intermediate point $\widetilde{\theta}$ is measurable}?
	
	\noindent Unfortunately, both answers are ``no" due to the ``deterministic in, deterministic out'' principle of the classical theorems. 
	These issues are crucial in statistical applications.
	The randomness of the likelihood function $\ell(\widehat{\theta}; X)$ and the estimator $\widehat{\theta}(X)$ poses challenges in rigorously applying Taylor-related theorems in statistical contexts, 
	potentially undermining subsequent probabilistic arguments concerning statistical properties, such as consistency (i.e., estimators converge in probability) or asymptotic normality (i.e., estimators converge in distribution). 
	
	To address this limitation,
	we propose a random version of the mean value theorem (and Taylor's theorem) that follows a ``\emph{random in, random out}'' principle:
	
	\textit{
		(Random mean value theorem) Suppose $\bm{X}$, $\bm{Y}$, and $\bm{Z}$ are random vectors and $f(\bm{Z}, \cdot)$ is a random function satisfies certain smoothness conditions.
		Then, there exists a random vector $\bm{\xi}$ such that $f\left(\bm{Z}, \bm{X} + \bm{Y} \right) = f \left(\bm{Z}, \bm{X}\right) + \nabla f \left(\bm{Z}, \bm{\xi}\right) \bm{Y}$, 
		where $\nabla f$ is the row vector of the gradient of $f$ (with respect to its second argument).
	}
	
	\noindent However, proving this measurable version of Taylor-related theorems is nontrivial.
	Elementary approaches, 
	such as a ``pointwise'' argument and a ``function composition'' argument, 
	fail to establish measurability, 
	as demonstrated by counterexamples constructed using a non-measurable set. 
	For ease of presentation, 
	we set $p = 1$, 
	assume function $f(Z; \cdot) = f(\cdot)$ is deterministic and only $X$ and $Y$ are random variables. 
	In this setting, 
	consider the ``pointwise'' argument: for every $\omega \in \Omega$, 
	$X(\omega)$ is a deterministic number. 
	The classical mean value theorem guarantees the existence of a deterministic number $\xi_{\omega}$ (which may not be unique), 
	located on the segment between $X(\omega)$ and $X(\omega) + Y(\omega)$,  
	such that $f\left(X(\omega) + Y(\omega)\right) = f\left(X(\omega)\right) + \nabla f \left(\xi_{\omega}\right) Y(\omega)$. 
	Even if we assume uniqueness for each $\omega$ and allow $\omega$ to vary over the entire $\Omega$, 
	the resulting pointwise-defined map $\xi: \Omega \to \mathbb{R}, \omega \mapsto \xi_{\omega}$ does not necessarily ensure measurability.
	A counterexample involves the indicator function on a non-measurable set: let $\Omega = \mathbb{R}$ and $N \subseteq \Omega$ be a non-measurable set. 
	Then the pointwise-defined indicator function $\mathbb{I}_{N}: \Omega \to \{0, 1\}$, which takes the value $1$ if $\omega \in N$ and $0$ otherwise, is not measurable.
	Another approach involves an implicit ``function composition'' argument: $\nabla f\left(\xi\right) = \left\{f\left(X + Y\right) - f \left(X\right)\right\} Y ^ {- 1}$.
	One might incorrectly argue that $\xi$ is measurable based on the following false claim: if both the function $g$ (e.g., $\nabla f$) and the composition $g(h)$ (e.g., $\left\{f\left(X + Y\right) - f \left(X\right)\right\} Y ^ {- 1}$) are measurable, then the function $h$ (e.g., $\xi$) is measurable. 
	Unfortunately, this claim is false.
	A counterexample is the function $h: [- 1, 1] \to \{- 1, 1\}$, 
	which takes the value $1$ on a non-measurable subset of $[- 1, 1]$ and $- 1$ otherwise.
	Let $g: [-1, 1] \to [0, 1], x \mapsto \vert x \vert$ be the absolute value function, which is continuous and therefore measurable.
	We find that $g(h)(x) \equiv 1$ is measurable, 
	while $g$ is measurable, but $h$ is not.
	
	Existing literature provides valuable insights, but does not directly resolve the randomness and the measurability issues.
	For example, \citet{BenichouGail1989} applied the implicit function theorem to define random vectors implicitly,
	and \citet{MasseyWhitt1993, Lin1994, Crescenzo1999} developed probabilistic versions of Taylor expansions under expectation.
	However, these approaches impose restrictions, such as requiring non-negative random variables with finite means.
	\citet{AliprantisBorder2006} established a univariate version of Taylor's theorem for a deterministic function and a random variable, demonstrating that a measurable assignment $\xi(\omega)$ exists.
	Despite these contributions, the literature remains largely fragmented.
	%While researchers with strong analytical backgrounds may be familiar with these results, a user-friendly presentation tailored to common statistical applications is not readily available. 
	%Consequently, the precise status of the theorem is often unclear, particularly to graduate students in statistics and related fields.
	
	In this paper,
	we propose the multivariate Taylor-related theorems for random functions and random variables and provide simplified proofs under the mild assumptions.
	The proposed results address the randomness and the measurability gaps in statistical applications, offering a rigorous foundation for many Taylor-based proofs in statistics.
	The rest of the paper is organized as follows. 
	Section~\ref{Sec:results} introduces the proposed Taylor's theorems in two scenarios: deterministic functions and random variables, and random functions and random variables.
	Section~\ref{Sec:applications} discusses statistical applications, presenting a statistical version of Taylor-related theorems for random log-likelihood functions and estimators.
	%Section~\ref{Sec:extensions} extends the main result to discontinuous functions. 
	Additional discussions are presented in Section~\ref{Sec:discussion}.
	Classical theorems, definitions, technical lemmas, and proofs of our main results are provided in the \href{Supplementary Material.pdf}{Supplementary Material}.

\section{Main Results} \label{Sec:results}

	We adopt the classical notation provided in \cite{Lang1993}.
	Suppose $\mathcal{E}$ and $\mathcal{F}$ are two Banach spaces. 
	Let $\mathcal{U} \subseteq \mathcal{E}$ be an open set, 
	and let $\bm{f}: \mathcal{U} \to \mathcal{F}$ be of class $\mathcal{C} ^ {m}$ for a given integer $m \geq 1$, 
	i.e., $\bm{f}$ and $\bm{f} ^ {(k)}$ are continuous on $\mathcal{U}$ for $k = 0, 1, \ldots, m$. 
	Here, $\bm{f} ^ {(k)}$ denotes the $k$-th order derivative of $\bm{f}$ for integer $k \geq 0$ (for convenience, we define $\bm{f} ^ {(0)} \triangleq \bm{f}$ and $0! \triangleq 1$).
	Furthermore, let $\bm{x} \in \mathcal{U}$ and $\bm{y} \in \mathcal{E}$ such that $\bm{x} + t\bm{y} \in \mathcal{U}$ for all $t \in [0, 1]$.
	Then, we have
	\begin{equation}
		\label{Eq:main}
		\bm{f} \left(\bm{x} + \bm{y}\right) = 
		\sum_{k = 0}^{m - 1} \frac{1}{k!} \bm{f} ^ {(k)} \left(\bm{x}\right) \bm{y} ^ {(k)} + \bm{R}_m, 
		\quad
		\bm{R}_m \triangleq \int_{0}^{1} \frac{(1 - t) ^ {m - 1}}{(m - 1)!} \bm{f} ^ {(m)}\left(\bm{x} + t \bm{y} \right) \bm{y} ^ {(m)} \, \mathrm{d} t,
	\end{equation}
	where $\bm{y} ^ {(k)}$ represents the $k$-tuple $\left(\bm{y}, \ldots, \bm{y}\right)$ for integer $k \geq 1$ \citep{Lang1993}, 
	and $\bm{R}_m$ is the remainder in integral form.
	The deterministic versions of Taylor's and the mean value theorems are formally provided in the \href{Supplementary Material.pdf}{Supplementary Material}.
	
	\subsection{Deterministic functions} \label{Subsec:df}
	
	We now present extensions of the classical Taylor-related theorems: retaining deterministic functions while replacing deterministic vectors with random ones.  
	Throughout the paper, let $\Sigma_{\mathcal{A}}$ and $\mathcal{B}_{\mathcal{A}}$ denote the $\sigma$-algebra and Borel $\sigma$-algebra of subsets of set $\mathcal{A}$, respectively.
	
	\begin{theorem}[Taylor's Theorems for Deterministic Functions and Random Vectors]
		\label{Thm:dfrv}
		Let $\left(\Omega, \Sigma_{\Omega}\right)$, 
		$\left(\mathcal{X}, \mathcal{B}_{\mathcal{X}}\right)$, 
		$\left(\mathcal{Y}, \mathcal{B}_{\mathcal{Y}}\right)$, 
		and $\left(\mathcal{U}, \mathcal{B}_{\mathcal{U}}\right)$ be measurable spaces, 
		where 
		$\mathcal{U} \subseteq \mathcal{E} = \mathbb{R} ^ {p}$, 
		$\mathcal{F} = \mathbb{R}$, 
		$\mathcal{X} \subseteq \mathcal{U}$, 
		$\mathcal{Y} \subseteq \mathcal{E}$, 
		and $m \geq 1$ is an integer.
		%%% 
		Let $f: \mathcal{U} \to \mathcal{F}$ be a deterministic function.
		Suppose the following three conditions hold:
		
		(D-1) $\mathcal{U}$ is an open set. 
		
		(D-2) $f(\cdot)$ is $m$-th continuously differentiable on $\mathcal{U}$. 
		
		(D-3) Let $\bm{X}: \left(\Omega, \Sigma_{\Omega}\right) \to \left(\mathcal{X}, \mathcal{B}_{\mathcal{X}}\right)$ and $\bm{Y}: \left(\Omega, \Sigma_{\Omega}\right) \to \left(\mathcal{Y}, \mathcal{B}_{\mathcal{Y}}\right)$ be $p$-dimensional random vectors such that $\left\{\bm{X}(\omega) + t \bm{Y}(\omega): t \in [0, 1] \right\} \subseteq \mathcal{U}$ for all $\omega \in \Omega$.
		
		Then, there exists a $p$-dimensional random vector $\bm{\xi}: \left(\Omega, \Sigma_{\Omega}\right) \to \left(\mathcal{U}, \mathcal{B}_{\mathcal{U}}\right)$ such that 
		\begin{equation*}
			f\left(\bm{X} + \bm{Y} \right) 
			= \sum_{k = 0}^{m - 1} 
			\frac{1}{k!} \left(\sum_{\ell = 1}^{p} Y_{\ell} \frac{\partial}{\partial u_{\ell}} \right) ^ {k} f\left(\bm{X}\right) + 
			\frac{1}{m!} \left(\sum_{\ell = 1}^{p} Y_{\ell} \frac{\partial}{\partial u_{\ell}}\right) ^ {m}
			f\left(\bm{\xi}\right), 
		\end{equation*}
		and $\bm{\xi}(\omega) \in \left\{\bm{X}(\omega) + t\bm{Y}(\omega): t \in [0, 1]\right\} \subseteq \mathcal{U}$ for each $\omega \in \Omega$, 
		where $u_{\ell}$ and $Y_{\ell}$ are the $\ell$-th components of $\bm{u}$ and $\bm{Y}$, and the operator is short for 
		\begin{equation*}
			\left(\sum_{\ell = 1}^{p} Y_{\ell} \frac{\partial}{\partial u_{\ell}} \right) ^ {k}
			= \sum_{i_1, \ldots, i_p \in \mathbb{N}: \sum_{\ell = 1} ^ {p} i_{\ell} = k}
			\frac{k!}{i_1! \cdots i_p!} Y_1 ^ {i_1} \cdots Y_p ^ {i_p}
			\frac{\partial ^ k}{\partial u_1 ^ {i_1} \cdots \partial u_p ^ {i_p}}, \quad k = 0, 1, \ldots, m - 1. 
		\end{equation*}
		
		In particular, for $m = 1$, the mean value theorem holds:  
		there exists a $p$-dimensional random vector $\bm{\xi}: \left(\Omega, \Sigma_{\Omega}\right) \to \left(\mathcal{U}, \mathcal{B}_{\mathcal{U}}\right)$ such that
		\begin{equation*}
			f\left(\bm{X} + \bm{Y}\right) - f\left(\bm{X}\right) = \nabla f \left(\bm{\xi} \right) \bm{Y},
		\end{equation*}
		and $\bm{\xi}(\omega) \in \left\{\bm{X}(\omega) + t\bm{Y}(\omega): t \in [0, 1]\right\} \subseteq \mathcal{U}$ for each $\omega \in \Omega$, 
		where $\nabla f\left(\bm{u}\right) \triangleq \left(\partial f(\bm{u})/ \partial u_1, \ldots, \partial f(\bm{u})/ \partial u_p \right)$ is the gradient (row) vector.
	\end{theorem}
	
	Theorem~\ref{Thm:dfrv} extends Taylor's expansion to deterministic functions evaluated at random vectors, 
	with the intermediate point also shown to be random.
	While this generalization involves replacing only deterministic vectors with random ones, 
	the proof is nontrivial, as previously discussed.
	A detailed proof is provided in the \href{Supplementary Material.pdf}{Supplementary Material}. 
	A direct application of Theorem~\ref{Thm:dfrv}, 
	the proof of the Delta method, is presented in Section~\ref{Subsec:dfrv}.
	
	\begin{remark}[Univariate, Multivariate, and Vector-Valued Cases]
		\label{Rem:pq}
		Theorem~\ref{Thm:dfrv} presents the multivariate version of extended Taylor-related theorems for $p \geq 1$.
		When $p = 1$, 
		it naturally reduces to the univariate case.
		Additionally, let $\mathcal{E} = \mathbb{R} ^ {p}$ and $\mathcal{F} = \mathbb{R} ^ {q}$, where $p \geq 1$ and $q > 1$ are integers.
		In this vector-valued setting, 
		The classical Taylor's Theorem can still be formulated in the integral form~\eqref{Eq:main}, 
		where $\bm{f}(\bm{u}) \triangleq \left(f_1(\bm{u}), \ldots, f_q(\bm{u})\right) ^ \top: \mathbb{R} ^ {p} \to \mathbb{R} ^ {q}$ and each component $f_r\left(\bm{u}\right) \triangleq f_r \left(u_1, \ldots, u_p\right): \mathbb{R} ^ {p} \to \mathbb{R}$ for $r = 1, \ldots, q$. 
		Higher-order derivatives are computed iteratively via the first-order derivative matrix, given by $\nabla \bm{f}(\bm{u}) = \left(\partial f_r(\bm{u}) / \partial u_{\ell}\right)_{r = 1, \ell = 1} ^ {q, p} \in \mathbb{R} ^ {q \times p}$. 
		However, unlike the scalar case, 
		there is no direct vector-valued analogue of the deterministic mean value theorem (and consequently, no such result in a measurable sense).
		It is because there does not exist a universal $\bm{\xi}$ lying between $\bm{x}$ and $\bm{y}$ (applicable simultaneously across all components of $\bm{f}$) such that $\bm{f}\left(\bm{x} + \bm{y}\right) - \bm{f}(\bm{x}) = \nabla \bm{f}\left(\bm{\xi}\right) \bm{y}$, as discussed in \citet{Tao2011, Tao2022}.
		This limitation has led to occasional misapplications in the statistical literature.
		A solution, which utilizes Taylor's theorem in its integral form, is outlined in \citet{FengWangHan2013}.
	\end{remark}
	
	\begin{remark}[Continuity of $f ^ {(m)}$]
		\label{Rem:continuity}
		Certain versions of the classical Taylor's theorem require only the existence of $f ^ {(m)}$ on $\mathcal{U}$ without assuming its continuity \citep{Rudin1976}.
		In contrast, Theorem~\ref{Thm:dfrv} imposes the continuity condition (D-2) on $f ^ {(m)}$.
		Later in the paper, we relax this continuity assumption.
	\end{remark}
	
	Building on Remark~\ref{Rem:continuity}, 
	we present the following extension of Theorem~\ref{Thm:dfrv}, 
	where $f ^ {(m)}$ is not continuous on $\mathcal{U}$.
	
	\begin{corollary}[Taylor's Theorems for Quasi-Continuous Functions with Closed Graphs]
		\label{Cor:qfrv}
		Suppose conditions (D1) and (D3) in Theorem~\ref{Thm:dfrv}, along with the following condition hold:
		
		(D-2') $f(\cdot)$ is $(m - 1)$-th continuously differentiable and $f ^ {(m)}(\cdot)$ is quasi-continuous with a closed graph on $\mathcal{U}$.
		
		Then, the conclusions in Theorem~\ref{Thm:dfrv} still hold.
	\end{corollary}
	
	A detailed proof is provided in the \href{Supplementary Material.pdf}{Supplementary Material}. 
	The definition of quasi-continuous functions is provided in the \href{Supplementary Material.pdf}{Supplementary Material}.
	Corollary~\ref{Cor:qfrv} ensures the measurability of intermediate points for quasi-continuous functions with closed graphs, even when they are discontinuous. 
	An example is the piecewise continuous function $f(x) = \begin{cases}
		1 / x, & x > 0 \\ 0, & x \leq 0
	\end{cases}$,
	which is discontinuous but is quasi-continuous with a closed graph \citep{DindosToma1997}. 
	
	\subsection{Random functions} \label{Subsec:rf}
	
	In this section, we continue to work with random vectors while replacing deterministic functions with random ones.
	
	\begin{theorem}[Taylor's Theorems for Random Functions and Random Vectors]
		\label{Thm:rfrv}
		Let $\left(\Omega, \Sigma_{\Omega}\right)$, 
		$\left(\mathcal{X}, \mathcal{B}_{\mathcal{X}}\right)$, 
		$\left(\mathcal{Y}, \mathcal{B}_{\mathcal{Y}}\right)$, 
		$\left(\mathcal{Z}, \mathcal{B}_{\mathcal{Z}}\right)$, 
		$\left(\mathcal{U}, \mathcal{B}_{\mathcal{U}}\right)$, 
		and $\left(\mathcal{F}, \mathcal{B}_{\mathcal{F}}\right)$ be measurable spaces, 
		where 
		$\mathcal{U} \subseteq \mathcal{E} = \mathbb{R} ^ {p}$, 
		$\mathcal{F} = \mathbb{R}$, 
		$\mathcal{X} \subseteq \mathcal{U}$, 
		$\mathcal{Y} \subseteq \mathcal{E}$, 
		$\mathcal{Z} \subseteq \mathbb{R} ^ {d}$, 
		and $m \geq 1$ is an integer. 
		%%%
		Let $\bm{Z}: \left(\Omega, \Sigma_{\Omega}\right) \to \left(\mathcal{Z}, \mathcal{B}_{\mathcal{Z}}\right)$ be a $d$-dimensional random vector and $f: \mathcal{Z} \times \mathcal{U} \to \mathcal{F}$ be a bivariate function.
		Thus,  
		$f(\bm{Z}, \bm{u})$ is a random function.
		%%%
		Suppose the following four conditions hold:
		
		(R-1) $\mathcal{U}$ is an open set.
		
		(R-2) For each $\bm{u} \in \mathcal{U}$, $f \left(\cdot, \bm{u}\right)$ is continuous in argument $\bm{z}$.
		
		(R-3) For each $\bm{z} \in \mathcal{Z}$, 
		$f\left(\bm{z}, \cdot\right)$ is $m$-th continuously differentiable in argument $\bm{u}$ on $\mathcal{U}$.
		
		(R-4) Let $\bm{X}: \left(\Omega, \Sigma_{\Omega}\right) \to \left(\mathcal{X}, \mathcal{B}_{\mathcal{X}}\right)$ and $\bm{Y}: \left(\Omega, \Sigma_{\Omega}\right) \to \left(\mathcal{Y}, \mathcal{B}_{\mathcal{Y}}\right)$ be $p$-dimensional random vectors such that $\left\{\bm{X}(\omega) + t \bm{Y}(\omega): t \in [0, 1] \right\} \subseteq \mathcal{U}$ for all $\omega \in \Omega$.
		
		Then, there exists a $p$-dimensional random vector $\bm{\xi}: \left(\Omega, \Sigma_{\Omega}\right) \to \left(\mathcal{U}, \mathcal{B}_{\mathcal{U}}\right)$ such that 
		\begin{equation*}
			f\left(\bm{Z}, \bm{X} + \bm{Y} \right) 
			= \sum_{k = 0}^{m - 1} 
			\frac{1}{k!} \left(\sum_{\ell = 1}^{p} Y_{\ell} \frac{\partial}{\partial u_{\ell}} \right) ^ {k} f\left(\bm{Z}, \bm{X}\right) + 
			\frac{1}{m!} \left(\sum_{\ell = 1}^{p} Y_{\ell} \frac{\partial}{\partial u_{\ell}}\right) ^ {m}
			f\left(\bm{Z}, \bm{\xi}\right),
		\end{equation*} 
		and $\bm{\xi}(\omega) \in \left\{\bm{X}(\omega) + t\bm{Y}(\omega): t \in [0, 1]\right\} \subseteq \mathcal{U}$ for each $\omega \in \Omega$, 
		where $u_{\ell}$ and $Y_{\ell}$ are the $\ell$-th components of $\bm{u}$ and $\bm{Y}$, and the operator is short for 
		\begin{equation*}
			\left(\sum_{\ell = 1}^{p} Y_{\ell} \frac{\partial}{\partial u_{\ell}} \right) ^ {k}
			= \sum_{i_1, \ldots, i_p \in \mathbb{N}: \sum_{\ell = 1} ^ {p} i_{\ell} = k}
			\frac{k!}{i_1! \cdots i_p!} Y_1 ^ {i_1} \cdots Y_p ^ {i_p}
			\frac{\partial ^ k}{\partial u_1 ^ {i_1} \cdots \partial u_p ^ {i_p}}, \quad k = 0, 1, \ldots, m - 1. 
		\end{equation*}
		
		In particular, for $m = 1$, the mean value theorem holds:  
		there exists a $p$-dimensional random vector $\bm{\xi}: \left(\Omega, \Sigma_{\Omega}\right) \to \left(\mathcal{U}, \mathcal{B}_{\mathcal{U}}\right)$ such that
		\begin{equation*}
			f\left(\bm{Z}, \bm{X} + \bm{Y}\right) - f\left(\bm{Z}, \bm{X}\right) = \nabla_{\bm{u}} f \left(\bm{Z}, \bm{\xi} \right) \bm{Y},
		\end{equation*}
		and $\bm{\xi}(\omega) \in \left\{\bm{X}(\omega) + t\bm{Y}(\omega): t \in [0, 1]\right\} \subseteq \mathcal{U}$ for each $\omega \in \Omega$, 
		where, with respect to the second argument, $\nabla_{\bm{u}} f\left(\bm{z}, \bm{u}\right) \triangleq \left(\partial f(\bm{z}, \bm{u})/ \partial u_1, \ldots, \partial f(\bm{z}, \bm{u})/ \partial u_p \right)$ is the gradient (row) vector.
	\end{theorem}
	
	A detailed proof is provided in the \href{Supplementary Material.pdf}{Supplementary Material}. 
	Theorem~\ref{Thm:rfrv} generalizes the deterministic functions in Theorem~\ref{Thm:dfrv} to a specific random setting, 
	where randomness arises from evaluating the functions at random vectors.
	Although Theorem~\ref{Thm:rfrv} is not directly applicable to most statistical problems, 
	we include it for readers with a particular interest in the result. 
	A statistically oriented version is presented in Theorem~\ref{Thm:likelihood_n}.
	
\section{Statistical Applications} \label{Sec:applications}
	
	The extended Taylor's theorems have broad applicability across various statistical domains, 
	particularly where Taylor-based methods are essential for understanding statistical results or proofs.
	In this section,
	we demonstrate how our results enhance comprehension of these topics and address potential technical gaps.
	
	\subsection{Deterministic functions and random vectors} \label{Subsec:dfrv}
	
	When statistical derivations involve deterministic functions and random variables, using Theorem~\ref{Thm:dfrv} enhances both rigor and clarity. 
	An example illustrating this improvement is the derivation of the Delta method.
	
	\textbf{Example (the Delta method)} \citep[Example 4.4]{Jiang2010}.
	Suppose $g(\cdot)$ is a continuously differentiable multivariate function defined on an open set. 
	Consider a sequence of random vectors $\left\{\bm{\xi}_n\right\}_{n = 1} ^ {\infty}$ and a random vector  $\bm{\eta}$ such that $a_n \left(\bm{\xi}_n - \bm{c}\right) \to \bm{\eta}$ in distribution, 
	where $\bm{c}$ is a constant vector and $\{a_n\}_{n = 1} ^ {\infty}$ is a positive sequence diverging to infinity.
	To establish the Delta method, 
	the classical mean value theorem is applied, yielding
	\begin{equation*}
		g\left(\bm{\xi}_n\right) = g(\bm{c}) + \nabla g \left(\bm{\zeta}_n\right) \left(\bm{\xi}_n - \bm{c}\right), 
	\end{equation*}
	where ``$\bm{\zeta}_n$ lies between $\bm{c}$ and $\bm{\xi}_n$''.
	However, 
	since $\bm{\xi}_n$ is random, 
	the classical mean value theorem does not formally guarantee the measurability of $\bm{\zeta}_n$, 
	rendering the argument incomplete in probabilistic terms.
	
	All conditions of Theorem~\ref{Thm:dfrv} are satisfied in this setting.
	Therefore, 
	we apply it to the deterministic function $g$, the random vector $\bm{\xi}_n$, and the deterministic vector $\bm{c}$.
	This establishes that $\bm{\zeta}_n$ is a \emph{random vector} lying between $\bm{\xi}_n$ and $\bm{c}$.
	As a result, 
	probabilistic statements such as $\nabla g (\bm{\zeta}_n) - \nabla g (\bm{c}) = o_P(1)$ become valid, 
	since $\bm{\zeta}_n$ is random here.
	This satisfies the conditions required for applying Slutsky's theorem in subsequent asymptotic analysis. \hfill $\blacksquare$

	\subsection{Random functions and random vectors} \label{Subsec:rfrv}

	One of the most important applications of our main results arises in maximum likelihood estimation, 
	which frequently involves Taylor expansions of random score functions.
	In this context, both functions and variables are stochastic.
	Before presenting these applications, 
	we briefly review likelihood-based parameter estimation, 
	following the notation of \citet{IbragimovHasMinskii1981}. 
	
	Let $\Omega$ be the set of elementary outcomes, 
	let $\mathcal{Z} \subseteq \mathbb{R} ^ {d}$ represent the set of observation values, 
	and let $\Theta \subseteq \mathbb{R} ^ {p}$ represent the parameter set. 
	Assume that $\left(\Omega, \Sigma_{\Omega}\right)$, $\left(\mathcal{Z}, \mathcal{B}_{\mathcal{Z}}\right)$, and $\left(\Theta, \mathcal{B}_{\Theta}\right)$ are measurable spaces.
	
	\emph{A single observation.} 
	Define a $d$-dimensional random vector $\bm{Z}: \left(\Omega, \Sigma_{\Omega}\right) \to \left(\mathcal{Z}, \mathcal{B}_{\mathcal{Z}}\right)$.
	Suppose its distribution belongs to the parametric family $\left\{P_{\theta}, \theta \in \Theta \right\}$, 
	possessing a dominating $\sigma$-finite measure $\mu$ (e.g., the counting measure for a discrete random vector or the Lebesgue measure for a continuous random vector).
	By the Radon-Nikodym theorem, 
	for each $\theta \in \Theta$, 
	there exists a non-negative density function $p(\bm{z}; \theta) \triangleq \mathrm{d} P_{\theta}(\bm{z}) / \mathrm{d} \mu(\bm{z})$ with respect to $\mu$. 
	We emphasize that, 
	by definition, the univariate mapping $p(\bm{z}; \theta): \mathcal{Z} \to \mathbb{R}$ is the \emph{density function}; 
	and switching the arguments, 
	the univariate mapping $p(\theta; \bm{z}) \triangleq p(\bm{z}; \theta): \Theta \to \mathbb{R}$ is the \emph{likelihood function} (for a single observation).
	We define a bivariate mapping 
	\begin{equation*}
		\ell(\bm{z}, \theta) \triangleq \log p(\bm{z}; \theta): \mathcal{Z} \times \Theta \to \mathbb{R},
	\end{equation*}
	where, for each $\bm{z} \in \mathcal{Z}$, 
	$\ell_{\bm{z}}(\theta) \triangleq \ell(\bm{z}, \theta)$ serves as the \emph{log-likelihood function} (for a single observation) in argument $\theta$.
	Replacing deterministic vector $\bm{z}$ with random vector $\bm{Z}$, 
	we obtain a \emph{random} function $\ell\left(\bm{Z}, \theta\right) = \log p(\bm{Z}; \theta)$.
	
	\emph{Statistics and estimators.}
	Suppose $\mathcal{T}_1$ and $\mathcal{T}_2$ are subsets of Euclidean spaces, 
	$\left(\mathcal{T}_1, \mathcal{B}_{\mathcal{T}_1}\right)$ and $\left(\mathcal{T}_2, \mathcal{B}_{\mathcal{T}_2}\right)$ are measurable spaces. 
	We define measurable functions $\bm{T}_1: \left(\mathcal{Z}, \mathcal{B}_{\mathcal{Z}}\right) \to \left(\mathcal{T}_1, \mathcal{B}_{\mathcal{T}_1}\right)$ and $\bm{T}_2: \left(\mathcal{Z}, \mathcal{B}_{\mathcal{Z}}\right) \to \left(\mathcal{T}_2, \mathcal{B}_{\mathcal{T}_2}\right)$ as statistics.
	Thus, the composite mappings $\widetilde{\bm{T}}_1 \triangleq \bm{T}_1(\bm{Z})$ and $\widetilde{\bm{T}}_2 \triangleq \bm{T}_2(\bm{Z})$ are estimators from $\left(\Omega, \Sigma_{\Omega}\right)$ to $\left(\mathcal{T}_1, \mathcal{B}_{\mathcal{T}_1}\right)$ and $\left(\mathcal{T}_2, \mathcal{B}_{\mathcal{T}_2}\right)$, respectively.
	
	\emph{$n$ independent copies.}
	Suppose $\bm{Z}_1, \ldots, \bm{Z}_n$ are $n$ independent copies of $d$-dimensional random vector $\bm{Z}: \left(\Omega, \Sigma_{\Omega}\right) \to \left(\mathcal{Z}, \mathcal{B}_{\mathcal{Z}}\right)$.
	We consider the joint log-likelihood function of $\bm{Z}_1, \ldots, \bm{Z}_n$ as below.
	We define $\mathcal{Z} ^ n$, $\mathcal{B}_{\mathcal{Z}} ^ n$, and $P_{\theta} ^ n$ as the $n$-products $\mathcal{Z} \times \cdots \times \mathcal{Z}$, $\mathcal{B}_{\mathcal{Z}} \times \cdots \times \mathcal{B}_{\mathcal{Z}}$, and $P_{\theta} \times \cdots \times P_{\theta}$, respectively.
	All the measures $P_{\theta} ^ n$ possess the densities
	\begin{equation*}
		p_n (\bm{z}_{1:n}; \theta) \triangleq \frac{\mathrm{d}P_{\theta} ^ n(\bm{z}_{1:n})}{\mathrm{d} \mu ^ n(\bm{z}_{1:n})} = p(\bm{z}_1; \theta) \times \cdots \times p(\bm{z}_n; \theta), 
		\quad \bm{z}_{1:n} \triangleq \left(\bm{z}_1 ^ \top, \ldots, \bm{z}_n ^ \top \right) ^ \top \in \mathbb{R} ^ {n d}, 
	\end{equation*}
	with respect to the $n$-product measure $\mu ^ n \triangleq \mu \times \cdots \times \mu$.
	Thus, we define a bivariate mapping
	\begin{equation*}
		\ell_n(\bm{z}_{1:n}, \theta) \triangleq \log p_n\left(\bm{z}_{1:n}; \theta \right): \mathcal{Z} ^ n \times \Theta \to \mathbb{R},
	\end{equation*}
	where, for each $\bm{z}_{1:n} \in \mathcal{Z} ^ n$, $\ell_{n, \bm{z}_{1:n}} (\theta) \triangleq \ell_n \left(\bm{z}_{1:n}, \theta \right)$ is the \emph{joint log-likelihood function} in argument $\theta$.
	By stacking $n$ observations into one,  
	we can define a $(nd)$-dimensional random vector $\bm{Z}_{1:n} \triangleq \left(\bm{Z}_1 ^ \top, \ldots, \bm{Z}_n ^ \top \right) ^ \top: \left(\Omega, \Sigma_{\Omega}\right) \to \left(\mathcal{Z} ^ n, \mathcal{B}_{\mathcal{Z}} ^ n\right)$ satisfying that its log-likelihood function (for the single $(nd)$-dimensional observation $\bm{Z}_{1:n}$) is equivalent to the joint log-likelihood function (for i.i.d. $d$-dimensional observations $\bm{Z}_1, \ldots, \bm{Z}_n$).
	Replacing deterministic vector $\bm{z}_{1:n}$ with random vector $\bm{Z}_{1:n}$, 
	we obtain a \emph{random} function $\ell_n\left(\bm{Z}_{1:n}, \theta\right) = \log p_n(\bm{Z}_{1:n}; \theta)$.
	
	With the definitions of $n$-copied samples and estimators, 
	we now adapt Theorem~\ref{Thm:rfrv} to a specialized version suited for random joint log-likelihood functions.
	
	\begin{theorem}[Taylor's Theorems for Random Joint Log-Likelihood Functions and Estimators]
		\label{Thm:likelihood_n}
		Let $\left(\Omega, \Sigma_{\Omega}\right)$, 
		$\left(\mathcal{T}_1, \mathcal{B}_{\mathcal{T}_1}\right)$, 
		$\left(\mathcal{T}_2, \mathcal{B}_{\mathcal{T}_2}\right)$, 
		$\left(\mathcal{Z}, \mathcal{B}_{\mathcal{Z}}\right)$, 
		$\left(\Theta, \mathcal{B}_{\Theta}\right)$, 
		and $\left(\mathcal{F}, \mathcal{B}_{\mathcal{F}}\right)$ be measurable spaces, 
		where 
		$\Theta \subseteq \mathcal{E} = \mathbb{R} ^ {p}$, 
		$\mathcal{F} = \mathbb{R}$, 
		$\mathcal{T}_1 \subseteq \Theta$, 
		$\mathcal{T}_2 \subseteq \mathcal{E}$, 
		$\mathcal{Z} \subseteq \mathbb{R} ^ {d}$, 
		and $m \geq 1$ is an integer. 
		%%%
		Let $\bm{Z}: \left(\Omega, \Sigma_{\Omega}\right) \to \left(\mathcal{Z}, \mathcal{B}_{\mathcal{Z}}\right)$ be a $d$-dimensional random vector with a non-negative density function $p(\bm{z}; \theta)$.
		%Let $\bm{Z}: \left(\Omega, \Sigma_{\Omega}\right) \to \left(\mathcal{Z}, \mathcal{B}_{\mathcal{Z}}\right)$ and $p(\bm{z}; \theta)$ be a $d$-dimensional random vector and its non-negative density function. 
		%Let $\ell(\bm{z}, \theta) \triangleq \log p (\bm{z}; \theta)$ denote a bivariate function. 
		%%%		
		Suppose $\bm{Z}_1, \ldots, \bm{Z}_n$ are $n$ independent copies of $\bm{Z}$.
		Thus, we let $\bm{Z}_{1:n} \triangleq \left(\bm{Z}_1 ^ \top, \ldots, \bm{Z}_n ^ \top \right) ^ \top: \left(\Omega, \Sigma_{\Omega}\right) \to \left(\mathcal{Z} ^ n, \mathcal{B}_{\mathcal{Z}} ^ n \right)$ be a $(nd)$-dimensional random vector with a non-negative density function $p_n(\bm{z}_{1:n}; \theta) = \prod_{i = 1}^{n} p \left(\bm{z}_i; \theta\right)$,
		%Thus, we define $\bm{Z}_{1:n} \triangleq \left(\bm{Z}_1 ^ \top, \ldots, \bm{Z}_n ^ \top \right) ^ \top: \left(\Omega, \Sigma_{\Omega}\right) \to \left(\mathcal{Z} ^ n, \mathcal{B}_{\mathcal{Z}} ^ n \right)$ as a $(nd)$-dimensional random vector and $p_n(\bm{z}_{1:n}; \theta) = \prod_{i = 1}^{n} p \left(\bm{z}_i; \theta\right)$ is its non-negative density function, 
		where $\bm{z}_{1:n}$ is the realization of $\bm{Z}_{1:n}$.
		Let $\ell_n \left(\bm{z}_{1:n}, \theta\right) \triangleq \log p_n(\bm{z}_{1:n}; \theta) = \sum_{i = 1}^{n} \log p \left(\bm{z}_i, \theta\right)$ denote a bivariate function. 
		Thus, 
		for all $\bm{z}_{1:n} \in \mathcal{Z} ^ n$, 
		$\ell_{n, \bm{z}_{1:n}}(\theta) \triangleq \ell_n \left(\bm{z}_{1:n}, \theta\right)$ is the joint log-likelihood function of $\bm{Z}_1, \ldots, \bm{Z}_n$.
		Replacing $\bm{z}_{1:n}$ with $\bm{Z}_{1:n}$,  
		$\ell_{n, \bm{Z}_{1:n}}(\theta) \triangleq \ell_n \left(\bm{Z}_{1:n}, \theta\right) \triangleq \sum_{i = 1}^{n} \log p (\bm{Z}_i, \theta)$ is a random function.
		%%%
		Suppose the following four conditions hold:
		
		(L-1) $\Theta$ is an open set.
		
		(L-2) For each $\theta \in \Theta$, $\ell_n \left(\cdot, \theta\right) \triangleq \log p_n(\cdot; \theta)$ is continuous in argument $\bm{z}_{1:n}$.
		
		(L-3) For each $\bm{z}_{1:n} \in \mathcal{Z} ^ n$, 
		$\ell_n \left(\bm{z}_{1:n}, \cdot\right) \triangleq \log p_n (\bm{z}_{1:n}; \cdot)$ is $m$-th continuously differentiable in argument $\theta$ on $\Theta$.
		
		(L-4) Let $\bm{T}_1: \left(\mathcal{Z} ^ n, \mathcal{B}_{\mathcal{Z}} ^ n \right) \to \left(\mathcal{T}_1, \mathcal{B}_{\mathcal{T}_1}\right)$ and $\bm{T}_2: \left(\mathcal{Z} ^ n, \mathcal{B}_{\mathcal{Z}} ^ n \right) \to \left(\mathcal{T}_2, \mathcal{B}_{\mathcal{T}_2}\right)$ be $p$-dimensional statistics such that $\left\{\widetilde{\bm{T}}_1(\omega) + t \widetilde{\bm{T}}_2(\omega): t \in [0, 1] \right\} \subseteq \Theta$ for all $\omega \in \Omega$, 
		where the estimators $\widetilde{\bm{T}}_1 \triangleq \bm{T}_1(\bm{Z}_{1:n}(\omega)): \Omega \to \Theta$ and $\widetilde{\bm{T}}_2 \triangleq \bm{T}_2(\bm{Z}_{1:n}(\omega)): \Omega \to \mathcal{E}$.
		
		Then, there exists a $p$-dimensional random vector $\bm{\xi}: \left(\Omega, \Sigma_{\Omega}\right) \to \left(\Theta, \mathcal{B}_{\Theta}\right)$ such that 
		\begin{equation*}
			\ell_n \left(\bm{Z}_{1:n}, \widetilde{\bm{T}}_1 + \widetilde{\bm{T}}_2 \right) 
			= \sum_{k = 0}^{m - 1} 
			\frac{1}{k!} \left(\sum_{\ell = 1}^{p} \widetilde{T}_{2, \ell} \frac{\partial}{\partial \theta_{\ell}} \right) ^ {k} \ell_n \left(\bm{Z}_{1:n}, \widetilde{\bm{T}}_1\right) + 
			\frac{1}{m!} \left(\sum_{\ell = 1}^{p} \widetilde{T}_{2, \ell} \frac{\partial}{\partial \theta_{\ell}}\right) ^ {m}
			\ell_n \left(\bm{Z}_{1:n}, \bm{\xi}\right),
		\end{equation*} 
		and $\bm{\xi}(\omega) \in \left\{\widetilde{\bm{T}}_1(\omega) + t \widetilde{\bm{T}}_2(\omega): t \in [0, 1]\right\} \subseteq \Theta$ for each $\omega \in \Omega$, 
		where $\theta_{\ell}$ and $\widetilde{T}_{2, \ell}$ are the $\ell$-th components of $\theta$ and $\widetilde{\bm{T}}_2$, and the operator is short for 
		\begin{equation*}
			\left(\sum_{\ell = 1}^{p} \widetilde{T}_{2, \ell} \frac{\partial}{\partial \theta_{\ell}} \right) ^ {k}
			= \sum_{i_1, \ldots, i_p \in \mathbb{N}: \sum_{\ell = 1} ^ {p} i_{\ell} = k}
			\frac{k!}{i_1! \cdots i_p!} \widetilde{T}_{2, 1} ^ {i_1} \cdots \widetilde{T}_{2, p} ^ {i_p}
			\frac{\partial ^ k}{\partial \theta_1 ^ {i_1} \cdots \partial \theta_p ^ {i_p}}, \quad k = 0, 1, \ldots, m - 1. 
		\end{equation*}
		
		In particular, for $m = 1$, the mean value theorem holds:  
		there exists a $p$-dimensional random vector $\bm{\xi}: \left(\Omega, \Sigma_{\Omega}\right) \to \left(\Theta, \mathcal{B}_{\Theta}\right)$ such that
		\begin{equation*}
			\ell_n \left(\bm{Z}_{1:n}, \widetilde{\bm{T}}_1 + \widetilde{\bm{T}}_2\right) - \ell_n \left(\bm{Z}_{1:n}, \widetilde{\bm{T}}_1\right) = \nabla_{\theta} \ell_n \left(\bm{Z}_{1:n}, \bm{\xi} \right) \widetilde{\bm{T}}_2,
		\end{equation*}
		and $\bm{\xi}(\omega) \in \left\{\widetilde{\bm{T}}_1(\omega) + t \widetilde{\bm{T}}_2(\omega): t \in [0, 1]\right\} \subseteq \Theta$ for each $\omega \in \Omega$,
		where, with respect to the second argument, $\nabla_{\theta} \ell_n \left(\bm{z}_{1:n}, \theta\right) \triangleq \left(\partial \ell_n (\bm{z}_{1:n}, \theta)/ \partial \theta_1, \ldots, \partial \ell_n (\bm{z}_{1:n}, \theta)/ \partial \theta_p \right)$ is the gradient (row) vector.
	\end{theorem}
	
	A detailed proof is provided in the \href{Supplementary Material.pdf}{Supplementary Material}. 
	Theorem~\ref{Thm:likelihood_n} presents a statistical version of Taylor's theorem that incorporates a random objective function, $n$-copied random samples, and estimators.
	Its conditions are generally mild, with the exception of Condition (L-4), which requires estimators.
	This may need to be explicitly verified in some cases.
	In the example that follows, 
	we illustrate how Theorem~\ref{Thm:likelihood_n} facilitates the analysis of classical maximum likelihood estimation.
	
	\textbf{Example (maximum likelihood estimation)} \citep[Chapter 18.16]{KendallStuart1961}.
	In the seminal monograph, 
	the asymptotic normality of the maximum likelihood estimator (MLE) is derived using the classical mean value theorem. 
	Specifically,  
	let $\bm{Z}_{1:n}$ and $\ell_{n}(\theta) \triangleq \ell_{n}\left(\bm{Z}_{1:n}, \theta\right)$ denote the random sample of size $n$ and the random version of the joint log-likelihood function for any fixed $\bm{Z}_{1:n}$, respectively.
	The classical mean value theorem is applied to the score function $\ell_n ^ {(1)}(\theta)$, evaluated at the MLE $\widehat{\theta}_n \triangleq \widehat{\theta}_n(\bm{Z}_{1:n})$, 
	around the true parameter $\theta_0$, 
	\begin{equation*}
		%\label{Eq:score}
		\ell_n ^ {(1)}\left(\widehat{\theta}_n\right)
		= \ell_n ^ {(1)}\left(\theta_0\right) + \left(\widehat{\theta}_n - \theta_0 \right)
		\ell_n ^ {(2)}\left(\theta_n ^ {\star}\right), 
	\end{equation*}
	where ``$\theta_n ^ {\star}$ is some value between $\widehat{\theta}_n$ and $\theta_0$''.
	This step, however, 
	overlooks the randomness of the score function $\ell_n ^ {(1)}$ and the MLE $\widehat{\theta}_n$, 
	leaving the argument incomplete.  
	
	We assume that 
	(1) the parameter space $\Theta$ is open, 
	(2) $\ell_n ^ {(1)}(\bm{z}_{1:n}, \theta)$ is continuous in argument $\bm{z}_{1:n}$ and (3) continuously differentiable in argument $\theta$, 
	and (4) $\bm{T}_1 = \theta_0$ and $\bm{T}_2 = \widehat{\theta}_n - \theta_0$ are measurable satisfying that $\left\{(1 - t)\theta_0 + t \widehat{\theta}_n(\bm{Z}_{1:n}(\omega)): t \in [0, 1]\right\} \subseteq \Theta$ for all $\omega \in \Omega$.
	Thus, all assumptions of Theorem~\ref{Thm:likelihood_n} are satisfied.
	We apply it to the random version of the score function $\ell_n ^ {(1)}$, 
	the MLE $\widehat{\theta}_n(\bm{Z}_{1:n})$, 
	and the deterministic parameter $\theta_0$, 
	yielding that there exists a \emph{random vector} $\theta_n ^ {\star}$ such that
	\begin{equation*}
		\ell_n ^ {(1)} \left(\bm{Z}_{1:n}, \widehat{\theta}_n(\bm{Z}_{1:n})\right) - \ell_n ^ {(1)} \left(\bm{Z}_{1:n}, \theta_0\right) = \ell_n ^ {(2)} \left(\bm{Z}_{1:n}, \theta ^ {\star} \right) \left(\widehat{\theta}_n(\bm{Z}_{1:n}) - \theta_0\right),
	\end{equation*}
	and $\theta_n ^ {\star}(\omega) \in \left\{(1 - t) \theta_0 + t \widehat{\theta}_n\left(\bm{Z}_{1:n}(\omega)\right): t \in [0, 1]\right\} \subseteq \Theta$ for each $\omega \in \Omega$.
	We note measurability of $\theta_n ^ {\star}$ is crucial in \citet{KendallStuart1961}'s derivation of asymptotic normality of $\widehat{\theta}_n$, 
	because,
	without it,  
	the subsequent analyses, e.g., 
	$\Pr
	\left\{\lim \limits_{n \to \infty} \ell_n ^ {(2)}\left(\theta_n ^ {\star}\right) = - \operatorname{E} \left\{\ell_n ^ {(1), 2}\left(\theta_0\right) \right\} \right\} = 1$,
	may not be reasonable within a probabilistic framework. \hfill $\blacksquare$
		
	In the next two examples, 
	we demonstrate how Theorem~\ref{Thm:likelihood_n} can aid in analyzing random functions beyond the classical likelihoods. 
	
	\textbf{Example ($M$-estimation)} \citep[Chapter 5.3]{vanderVaart1998}.
	Let $\bm{Z}_{1:n}$, $\widehat{\theta}_n \triangleq \widehat{\theta}_n(\bm{Z}_{1:n})$, and $\Psi_n(\theta) \triangleq \Psi_n(\bm{Z}_{1:n}, \theta) = 0$ denote the random sample, the $M$-estimator, and a system of random equations (e.g., the average of score equations in maximum likelihood estimation) for any fixed $\bm{Z}_{1:n}$.
	In providing the asymptotic normality of $M$-estimators, 
	\citet{vanderVaart1998} expands $\Psi_n(\widehat{\theta}_n)$ using a classical version of the Taylor series around the true value $\theta_0$ and obtains that 
	\begin{equation*}
		0 = \Psi_n(\widehat{\theta}_n) = \Psi_n(\theta_0) + \left(\widehat{\theta}_n - \theta_0\right) \Psi_n ^ {(1)}(\theta_0)
		+ \frac{1}{2} \left(\widehat{\theta}_n - \theta_0 \right) ^ 2 \Psi_n ^ {(2)}\left(\widetilde{\theta}_n\right),
	\end{equation*} 
	where ``$\widetilde{\theta}_n$ is a point between $\widehat{\theta}_n$ and $\theta_0$''.
	Obviously, the classical Taylor's theorem fails due to a lack of consideration of the randomness of $\Psi_n$ and $\widehat{\theta}_n$.
	
	Under assumptions that (1) the parameter space is an open set, 
	(2) $\Psi_n(\bm{z}_{1:n}, \theta)$ is continuous in argument $\bm{z}_{1:n}$ and (3) twice continuously differentiable in argument $\theta$, 
	and (4) $\bm{T}_1 = \theta_0$ and $\bm{T}_2 = \widehat{\theta}_n - \theta_0$ are measurable satisfying that $\left\{(1 - t)\theta_0 + t \widehat{\theta}_n(\bm{Z}_{1:n}(\omega)): t \in [0, 1]\right\} \subseteq \Theta$ for all $\omega \in \Omega$.
	All assumptions of Theorem~\ref{Thm:likelihood_n} are satisfied.
	Applying Theorem~\ref{Thm:likelihood_n}, 
	we still obtain the above expansion while $\widetilde{\theta}_n$ is a \emph{random variable} between $\widehat{\theta}_n$ and $\theta_0$.
	Thus, it is reasonable to verify that $\Psi_n ^ {(2)}(\widetilde{\theta}_n) = O_P(1)$ in subsequent analysis for the asymptotic normality of $M$-estimators. \hfill $\blacksquare$

	\textbf{Example (profile estimation)} \citep[Chapter 19.3]{Kosorok2008}.
	Let $\left\{P_{\theta, \eta}: \theta \in \Theta, \eta \in \mathcal{H}\right\}$ denote a family of semiparametric models, 
	where $\Theta \subseteq \mathbb{R} ^ {p}$ is a finite-dimensional set and $\mathcal{H}$ is an infinite-dimensional set.
	Let $\bm{Z}_{1:n}$ and $\ell_n \left(\theta, \eta\right) \triangleq \ell_n \left(\bm{Z}_{1:n}, \theta, \eta \right)$ denote the random sample of size $n$ and the random version of the joint log-likelihood function for any fixed $\bm{Z}_{1:n}$.
	Additionally, define $\theta_0$ and $\eta_0$ as the true parameters.
	%Additionally, we define the random version of the profile log-likelihood function $p\ell_n (\theta) \triangleq \sup \limits_{\eta \in \mathcal{H}} \ell_n (\theta, \eta)$.
	
	Following the \citet{Kosorok2008}'s notation,
	let $\widehat{\eta}_{\theta_0} \triangleq \argmax \limits_{\eta \in \mathcal{H}} \ell_n \left(\theta_0, \eta \right)$, 
	let $\{\widetilde{\theta}_n\}_{n = 1} ^ {\infty}$ denote a sequence of estimators converging to the true value $\theta_0$ in probability, 
	and let $\widetilde{\psi} \triangleq \left(\theta_0, \widehat{\eta}_{\theta_0}\right)$.
	We assume there exists a map $t \mapsto \eta_{t}(\widetilde{\psi}) \triangleq \eta_{\widetilde{\psi}, t}$ from the neighborhood of $\theta_0$ into the parameter set for $\widehat{\eta}_{\theta_0}$, 
	satisfying that $\eta_{\widetilde{\psi}, t} \in \widehat{\mathcal{H}}$ if $\left \Vert t - \theta_0 \right \Vert$ is sufficiently small, and 
	$\eta_{\widetilde{\psi}, t} = \widehat{\eta}_{\theta_0}$ if $\widetilde{\psi} = \left(\theta_0, \widehat{\eta}_{\theta_0}\right) \in \Theta \times \widehat{\mathcal{H}}$, 
	where $\widehat{\mathcal{H}}$ is a suitable enlargement of $\mathcal{H}$.
	Thus,
	we rewrite $\eta_{\widetilde{\psi}, \theta_0} \triangleq \widehat{\eta}_{\theta_0}$ using $t = \theta_0$ and define $\eta_{\widetilde{\psi}, \widetilde{\theta}_n}$ using $t = \widetilde{\theta}_n$ (which is sufficiently close to $\theta_0$).
	Thus, \citet{Kosorok2008} applies the classical Taylor theorem and obtains that 
	\begin{equation*}
		\ell_n \left(\widetilde{\theta}_n, \eta_{\widetilde{\psi}, \widetilde{\theta}_n}\right)
		= 
		\ell_n \left(\theta_0, \eta_{\widetilde{\psi}, \theta_0}\right)
		+ \left(\widetilde{\theta}_n - \theta_0\right) ^ \top 
		\ell_n ^ {(1)} \left(\theta_0, \eta_{\widetilde{\psi}, \theta_0}\right)
		+ \frac{1}{2} \left(\widetilde{\theta}_n - \theta_0\right) ^ \top 
		\ell_n ^ {(2)} \left(\widetilde{t}_n, \eta_{\widetilde{\psi}, \widetilde{t}_n}\right)
		\left(\widetilde{\theta}_n - \theta_0\right),
	\end{equation*}
	for ``some $\widetilde{t}_n$ on the line segment between $\widetilde{\theta}_n$ and $\theta_0$".
	Unfortunately, the conclusion may be incomplete because the randomness of the function $\ell_{n}$ and the estimators $\widetilde{\theta}_n$, $\eta_{\widetilde{\psi}, \widetilde{\theta}_n}$, and $\eta_{\widetilde{\psi}, \theta_0}$ violates the assumptions in the classical Taylor theorem. 
	
	Given $\widetilde{\psi} \triangleq \widetilde{\psi}(\bm{Z}_{1:n}) \triangleq \left(\theta_0, \widehat{\eta}_{\theta_0}(\bm{Z}_{1:n})\right)$ and the map $t \mapsto \eta_{\widetilde{\psi}, t} \triangleq \eta_{\widetilde{\psi}(\bm{Z}_{1:n}), t}$, we rewrite
	\begin{equation*}
		\ell_n \left(\bm{Z}_{1:n}, \theta, \eta_{\widetilde{\psi}(\bm{Z}_{1:n}), \theta} \right)
		\triangleq 
		\ell_{\widetilde{\psi}, n} \left(\bm{Z}_{1:n}, \theta \right)
	\end{equation*}
	for $\theta$ sufficiently close to $\theta_0$.
	Suppose the following assumptions hold:
	(1) the parameter space $\Theta$ is open, 
	(2) $\ell_{\widetilde{\psi}, n} \left(\bm{z}_{1:n}, \theta \right)$ is continuous in argument $\bm{z}_{1:n}$ 
	and (3) is twice continuously differentiable in argument $\theta$, 
	and (4) $\bm{T}_1 = \theta_0$ and $\bm{T}_2 = \widetilde{\theta}_n - \theta_0$ are measurable satisfying that $\left\{(1 - t)\theta_0 + t \widetilde{\theta}_n(\bm{Z}_{1:n}(\omega)): t \in [0, 1]\right\} \subseteq \Theta$ for all $\omega \in \Theta$.
	Applying Theorem~\ref{Thm:likelihood_n}, 
	we obtain that 
	\begin{equation*}
		\ell_{\widetilde{\psi}, n} \left(\widetilde{\theta}_n\right)
		= 
		\ell_{\widetilde{\psi}, n} \left(\theta_0\right)
		+ \left(\widetilde{\theta}_n - \theta_0\right) ^ \top 
		\ell_{\widetilde{\psi}, n} ^ {(1)} \left(\theta_0\right)
		+ \frac{1}{2} \left(\widetilde{\theta}_n - \theta_0\right) ^ \top 
		\ell_{\widetilde{\psi}, n} ^ {(2)} \left(\widetilde{t}_n\right)
		\left(\widetilde{\theta}_n - \theta_0\right),
	\end{equation*}
	where $\widetilde{t}_n$ is a \emph{random vector} between $\widetilde{\theta}_n$ and $\theta$. 
	This implies that $n ^ {- 1} \ell_{\widetilde{\psi}, n} ^ {(2)} \left(\widetilde{t}_n\right) = n ^ {- 1} \ell_n ^ {(2)} (\widetilde{t}_n, \eta_{\widetilde{\psi}, \widetilde{t}_n}) \to P_0 \ell^{(2)}\left(\theta_0, \eta_{0} \right)$ in probability,
	where $\ell(\theta, \eta)$ denotes the random version of the log-likelihood function for a single observation,
	and $P_0$ is the probability measure at the true value. \hfill $\blacksquare$
	
	Analogous modifications help resolve arguments concerning the limiting behavior of MLEs in both the single-parameter case \citep{CoxHinkley1974, BhattacharyaLinPatrangenaru2016} and the general multi-parameter case \citep{KendallStuart1961, Jiang2010}.
	Additionally, 
	by applying Theorem~\ref{Thm:likelihood_n}, 
	we can address measurability gaps in the estimation of normal probabilities \citep[Chapter 4.2]{Lehmann2004}, 
	in the derivation of the Cram\'{e}r-Rao information inequality \citep[Chapter 4.3]{BhattacharyaLinPatrangenaru2016}, 
	and in the estimation procedure of the jackknife variance \citep[Chapter 5.5]{Shao2003}.
	More broadly, 
	our results extend to various areas, including the asymptotic properties of likelihood ratio statistics \citep[Chapter 16.2]{vanderVaart1998}, the analysis of convergence rates for penalized $M$-estimators \citep[Chapter 21.5]{Kosorok2008}.
			
\section{Discussion} \label{Sec:discussion}
	
	We have developed statistical extensions of Taylor's theorems and the mean value theorems,
	specifically for settings involving both random functions and random variables.
	We rigorously establish the measurable version of Taylor's theorems, 
	explicitly accounting for the randomness inherent in the functions.
	Extending beyond classical results, our theorems guarantee the existence of a random intermediate point between two given random vectors.
	Additionally, 
	we offer a sufficient condition for measurability: 
	in practice, continuity of the highest-order derivatives is typically sufficient, while theoretically, 
	this can be relaxed to quasi-continuity and closed graph conditions.
	We adapt the main results to a statistical setting, providing a formulation that is readily applicable to statistical problems.
	Our work represents a refined technical advancement that enhances clarity and simplifies statistical reasoning, particularly in the study of the asymptotic behavior of parametric, semiparametric, and nonparametric estimators. These results contribute to making statistical arguments more concise and accessible..

	\bigskip
	\begin{center}
		{\large\bf Supplementary Material}
	\end{center}
	
	The supplementary material contains the classical Taylor's theorems, additional definitions, technical lemmas, and proofs of our main results.

	\bigskip
	\begin{center}
		{\large\bf Acknowledgments}
	\end{center}
	
	We appreciate the time and effort that the editor, the associate editor, and the reviewers have dedicated to providing their valuable feedback on our manuscript.

	\bigskip
	\begin{center}
		{\large\bf Conflict of Interest Statement}
	\end{center}
	
	No potential competing interest was reported by the authors.

	\bigskip
	\begin{center}
		{\large\bf Funding}
	\end{center}
	
	This work was partially supported by the National Heart, Lung, and Blood Institute of the National Institutes of Health under Grant Number 1R01HL175410.

	\bibliographystyle{biom}	
	\bibliography{paper_ref_TaylorMVT}

\end{document}